%% file: main.tex
\RequirePackage{amsmath}
\documentclass[runningheads,a4paper]{llncs}
\usepackage{graphicx}
\usepackage{hyperref}
\usepackage{todonotes}
\usepackage{mathtools}
\usepackage{lineno}
\usepackage{amssymb}
\usepackage{enumitem}
\usepackage{cleveref}
\usepackage{listings}
\usepackage{tikz}
\usepackage{subfig}
\usetikzlibrary{automata,positioning,arrows,shapes,fit}

\usepackage{pgfplots}
\pgfplotsset{width=10cm,compat=1.9}

\newcommand{\veryshortarrow}[1][3pt]
{\mathrel
 {%
  \hbox{\usefont{U}{lasy}{m}{n}\symbol{40}}
  \mkern-4mu
  \vcenter{\hbox{\rule[-.2pt]{#1}{.4pt}}}%
 }
}

\makeatletter

\setbox0\hbox{$\xdef\scriptratio{\strip@pt\dimexpr
    \numexpr(\sf@size*65536)/\f@size sp}$}

\newcommand{\scriptveryshortarrow}[1][3pt]{{%
    \vcenter{\hbox{\rule[\scriptratio\dimexpr-.2pt\relax]
               {\scriptratio\dimexpr#1\relax}{\scriptratio\dimexpr.4pt\relax}}}%
   \mkern-4mu\hbox{\let\f@size\sf@size\usefont{U}{lasy}{m}{n}\symbol{41}}}}

\newcommand{\vgets}{\veryshortarrow}
\newcommand{\src}{\mathit{src}}
\newcommand{\trg}{\mathit{trg}}

\makeatother

\begin{document}
\title{Inform: From Compartmental Models to Stochastic Bounded Counter
Machines\thanks{Work supported by the Flemish inter-university (iBOF) ``DESCARTES'' project}}
\titlerunning{Compartmental Models as Stochastic Counter Machines}
%
\author{Tim Leys
\and
Guillermo A. P\'erez
}
\authorrunning{T. Leys and G. A. P\'erez}
%
\institute{University of Antwerp -- Flanders Make, Antwerpen, Belgium\\
\email{\{tim.leys, guillermo.perez\}@uantwerpen.be}}

\maketitle             
\begin{abstract}
Compartmental models are used in epidemiology to capture the evolution of infectious diseases such as COVID-19 in a population by assigning members of it to compartments with
labels such as susceptible, infected, and recovered. In a stochastic
compartmental model the flow of individuals between compartments is determined
probabilistically. We establish that certain stochastic compartment models
can be encoded as probabilistic counter machines where the configurations are
bounded. Based on the latter, we obtain simple descriptions of the models in
the PRISM language. This enables the analysis of such compartmental models via
probabilistic model checkers. Finally, we report on experimental results where
we analyze results from a Belgian COVID-19 model using a probabilistic
model checkers.

\end{abstract}

\input{sections/Introduction}
\input{sections/Preliminaries}

\input{sections/CompartmentToSCM}

\input{sections/Experiments}

\input{sections/conclusion}


\bibliographystyle{splncs04}
\bibliography{bibliography}

\newpage
\appendix
\input{sections/appendix}

\end{document}

%% file: sections/Introduction.tex
\section{Introduction}
Compartmental models are frequently used for studying the evolution of
infectious diseases and epidemics. In these models, the population is divided
into compartments and rates describe how people move between these
compartments. A popular and simple compartmental model for epidemiology is the
SIR model \cite{keeling2008modeling}, where a population is divided into three
groups: Susceptible, Infectious, and Recovered. People in the
susceptible group have not been infected by the pathogen and are hence
vulnerable to becoming infected by infectious people. Infectious people are
currently experiencing symptoms and can infect susceptible people they come
into contact with. Recovered people are removed from the system
because they are either recovered and immune, or have died.

Several variations of the simple compartmental models described above exist. These arise from changing the set of compartments into which the population is partitioned, or from considering deterministic dynamics (usually given as ordinary differential equations) or stochastic ones. In this work, we focus on stochastic compartmental models where the rates are given as binomial distributions (cf. chain binomial \cite{bailey1975mathematical}). For intuition, it is well known that such stochastic compartmental models define a Markov chain whose states correspond to the possible partitions of the population into the compartments considered. (In the sequel, we give one concrete encodings of these models into Markov chains.)

Compartmental models in general, mimic the way a pathogen spreads and are therefore often used to forecast or simulate possible future scenarios during an epidemic~\cite{holmdahl2020wrong}. For instance, a stochastic compartmental model was developed to capture the evolution of the early phase of the Belgian COVID-19 epidemic \cite{ABRAMS2021100449}. In the same work, the model was used to estimate the (future) impact of interventions such as social distancing and quarantine.
Due to the level of abstraction of (stochastic) compartmental models, some interesting mismatches arise with respect to the original (real-world) system. For instance, it is usually possible (though highly unlikely, given how the rates are defined) that all susceptible individuals instantaneously interact with some infectious individual and thus become infected themselves ``in one shot''. Additionally, for models with more compartments than the three mentioned earlier (susceptible, infectious, and recovered), the population is not guaranteed to remain constant. (Here, we mean that despite the model not explicitly considering birth or death, its semantics result in the population changing over time.) This is, for instance, the case in the stochastic model from~\cite{ABRAMS2021100449} and is in contrast to the fact that the simplest SIR model (without birth and death) does have a constant population. 

We focus on three questions regarding stochastic compartmental models. The first two can be loosely understood as validity or interpretability questions: How likely is it that the population is not constant along a simulation of the model; or that a large portion of the population moves from one compartment to another, in one shot?
The third question can instead be understood as a verification one: We are interested in determining the expected time before the end of the epidemic.
%
We see this as a verification problem since the model can incorporate a given policy which corresponds to interventions (by the local government) and affects the dynamics (read, the rates) of the model.



\paragraph*{Contributions}
In this work, we give a simple encoding of stochastic compartmental models
into Markov chains that is easy to implement in the PRISM
language~\cite{KNP11}. Then, we leverage this encoding to approach our guiding
questions above using the probabilistic model checkers
STORM~\cite{DBLP:journals/sttt/HenselJKQV22} and
Modest~\cite{DBLP:conf/tacas/HartmannsH14}. The encoding is fully automated
and implemented in our open-source: Inform.


\paragraph*{Related work}
There are 
formal-methods approaches for the analysis of models from biology or chemistry (see, e.g.~\cite{DBLP:journals/tcs/CiocchettaH09,DBLP:conf/birthday/BackenkohlerBW22}). 
For epidemiology, the most closely related work we have found is that of Wolf and co-authors, and in particular~\cite{DBLP:conf/qest/GrossmannBW20,DBLP:conf/hybrid/GrossmannBW21}. Nevertheless, the former focuses on deterministic compartmental models and the latter on branching processes instead of (stochastic) compartmental models.

%% file: sections/Preliminaries.tex
\section{Preliminaries}
For a set $\mathbb{S} \subseteq \mathbb{R}$ of numbers, we denote its nonnegative elements as: $\mathbb{S}_{>0} = \mathbb{S} \cap [0, +\infty)$.
We denote the cardinality of a set $S = \{ s_1, s_2, \dots, s_n\}$ as $|S|=n$.

We define a probability distribution over domain $\mathcal{X}$ as a function $d:\mathcal{X} \to [0, 1]$ s.t. $\sum_{x\in \mathcal{X}} d(x) = 1$. In this paper, all $\mathcal{X}$  are countable. We denote with $\mathrm{Dist}(\mathcal{X})$ the set of all distributions over $\mathcal{X}$.

Let $n\in\mathbb{N}$ and $p \in [0,1]$. The binomial distribution for $n$ and
$p$, written $B(n,p)$, is the discrete probability distribution of the number of successes of $n$
independent Bernoulli trials (or coin flips). The probability of success for
each trial is $p$. We write the probability of $k$ successes as $\Pr(k;n, p)$.

We denote the $i$-th element in an $n$-dimensional vector $\vec{v} \in
\mathbb{R}^n$ as $v_i$. We also define the dimension of a vector $\vec{v}$ as
$\mathrm{dim}(\vec{v})$. Finally, we define the $1$-norm of a vector $\vec{v}$
as $\lVert \vec{v} \rVert_{1} = \sum_{i = 1}^{\mathrm{dim}(\vec{v})} |v_i|$.

\subsection{Stochastic Compartmental Models }
A compartmental model is formally defined as a directed graph with labels $(V, E, \ell, pop)$, where the vertices $V$ represent the compartments, the egdes $E$ represent how individuals can change from one compartment to another, and $pop \in \mathbb{N}$ is the population size.
Finally, $\ell:E\rightarrow \mathbb{Q}_{>0} \cup (\mathbb{Q}_{>0})^{V}$ is a function that assigns values to the edges. These values represent the rates at which the partitioning of the population over the compartments will evolve over time.

For the remainder of this section, we fix a graph $\mathcal{G} = (V, E, \ell, pop)$ with compartments $V= \{c_1, c_2, \dots, c_n\}$ and edges $E = \{e_1, e_2, \dots, e_m\}$. We make the assumption that $\mathcal{G}$ has no self-loops, i.e. $(u,v) \in E$ implies $u \neq v$.



\paragraph{Compartment Flows and Time.} 
A compartmental model describes how a system (of individuals) evolves over
time. Time is represented as discrete time-steps and the system evolves in the
sense that individuals can move from their compartment to another compartment.
How individuals can move is captured by the edges of the graph $\mathcal{G}$.
If there is an edge $e = (c_x,c_y)$, at each time-step every individual in
$c_x$ can move to $c_y$. We define $\src(e) = c_{x}$ and $\trg(e)= c_{y}$.

We say any vector $(r_1, r_2, \dots, r_m)\in \mathbb{N}^m$ is a flow realization over $\mathcal{G}$:
it encodes the amount of people that effectively follow each edge during a single time step. 

\paragraph{System States and Transitions.}
We define the state of the system as a vector $\vec{s} = (s_1, s_2, \dots,
s_n)\in \mathbb{N}^n$ s.t. $\lVert \vec{s} \rVert_1 = pop$, where $s_i$ is the number of individuals in compartment $c_i$ for all $0 <i \leq n$. Let $\vec{s}$ and $\vec{s'}$ be states, we write $\vec{s}  \rightarrow \vec{s'}$ to denote that the state $\vec{s}$ transitions to $\vec{s'}$ in a single time-step. Formally, we write $\vec{s}  \rightarrow \vec{s'}$ iff there exists a flow realization $\vec{r}$ such, that for all $0< i \leq n$: 
\[
    s_i' - s_i = \sum_{\substack{0 < k \leq m,\\\trg(e_k) = c_i}} r_k - \sum_{\substack{0< j \leq m,\\\src(e_j) = c_i}} r_j;
    \text{ and }
    \sum_{\substack{0 < j \leq m,\\\src(e_j) = c_i}} r_j \leq s_i.
\]
The latter is necessary, but not sufficient for population to be preserved.

All realizations witnessing $\vec{s}\rightarrow\vec{s'}$ are solutions of the system of Diophantine equations $\vec{A}\vec{x} = \vec{s'} - \vec{s} \land -\vec{I}\vec{x} \leq 0 \land \vec{B}\vec{x} \leq \vec{s}$, where $\vec{A}$ and $\vec{B}$ are $n \times m$ matrices. Further, $A_{ij}= -1$ if $\src(e_j) = c_i$, $A_{ij}= 1$ if $\trg(e_j) = c_i$, and $A_{ij} = 0$ otherwise; and $B_{ij} = 1$ if $\src(e_j) = c_i$, and $B_{ij} = 0$ otherwise.

\paragraph{Transition Probabilities.}
Consider two states $\vec{s}$ and $\vec{s'}$. We discussed how individuals change
compartments, but not at what rate. This rate is given as a binomial
distribution where the success probability is determined by the $\ell$ function.
Let $\vec{r}$ be a witness of $\vec{s} \rightarrow \vec{s'}$ and suppose $r_i
> 0$ for a unique index $1 \leq i \leq m$.
If $\ell(e_i) \in \mathbb{Q}$, the actual success probability for the flow
$\vec{r}$ is $1-\exp(-h\ell(e_i))$
where $h \in
\mathbb{R}_{>0}$ is a constant that represents\footnote{This model is
actually a discretized-time stochastic approximation of the classical
compartmental models from epidemiology. We refer the interested reader
to~\cite{alarcon2023computation}, and references therein, for a slower
introduction to the model and how it approximates the more classical
epidemiological models.} the length of a
time-step~\cite{bailey1975mathematical}. Note that $\ell$ can also map to a
vector  $\vec{v} \in \mathbb{Q}^V$, indexed by the compartments $c_k \in V$. To avoid double indexing, we write $v_k$ instead of $v_{c_k}$. In this case, the
probability is $1-\prod_{0 < k \leq n}\exp(-h s_k  v_k)$. Hence, 
the probability that $r_i$ individuals follow $e_i$ is:
\[
  \Pr_{i}(\vec{s},\vec{r}) = 
  \begin{cases}
    \Pr(r_i; s_j, 1-\prod_{0 < k \leq n}\exp(-h s_k  v_k)) & \text{if } \ell(e_i) = \vec{v} \in \mathbb{Q}^V\\
    \Pr(r_i; s_j, 1-\exp(-h v)) & \text{if } \ell(e_i) = v \in \mathbb{Q}
  \end{cases}
\]
where $c_j = \src(e_i)$. We generalize this to flows $\vec{r}$ with multiple positive:
\(
  \Pr(\vec{s},\vec{r}) = \prod_{i=1}^m \Pr_i(\vec{s},
  \vec{r})
\)
where $\Pr_i(\cdot, \cdot) = 1$ if $r_i = 0$. Finally, we define the
probability of the transition $\vec{s} \rightarrow \vec{s'}$ as:
\(
  \Pr(\vec{s} \rightarrow \vec{s'}) = \sum_{\vec{r} \text{ witnessing }
                                                      \vec{s}\rightarrow
                                                 \vec{s'}} \Pr(\vec{s},
                                                 \vec{r}).
\)

\paragraph{Reaching Probabilities.}
A trajectory is a sequence $\tau = \vec{s}^{(0)} \dots \vec{s}^{(t)}$ of states such
that $\vec{s}^{(i)} \rightarrow \vec{s}^{(i+1)}$ for all $0 \leq i < t$. The
probability of the trajectory is the product of the probabilities of its
transitions: $\Pr(\tau) = \prod_{i=0}^{t-1} \Pr(\vec{s}^{(i)} \rightarrow
\vec{s}^{(i+1)})$, where the empty product is assumed to be $1$.
We write $\vec{s} \xrightarrow{*} \vec{s'}$ for the set of
all trajectories that start with $\vec{s}$ and such that $\vec{s'}$ appears
only at the end of the sequence. Finally,
we write $\Pr(\vec{s} \xrightarrow{*} \vec{s'})$ for the probability of
eventually reaching $\vec{s'}$ from $\vec{s}$. Formally, 
\(
  \Pr(\vec{s} \xrightarrow{*} \vec{s'}) = \sum_{\tau \in \vec{s} \xrightarrow{*}
  \vec{s'}} \Pr(\tau).
\)

\subsection{Discrete Time Markov Chains}
A discrete-time Markov chain is a tuple $\mathcal{M} = (S, P, \iota_{\mathit{init}},w)$, where $S$ is a countable, nonempty set of states, $P:S \to \mathrm{Dist}(S)$ is the transition probability function, $\iota_{\mathit{init}} \in \mathrm{Dist}(S)$ is the initial distribution, and $w : S \times S \to \mathbb{R}$ is a weight function. We say a state $s\in S$ is a possible initial state whenever $\iota_{\mathit{init}}(s) > 0$. In a similar way, all states $s' \in S$ s.t. $P(s)(s')>0$ are considered possible successor states of $s$. We say that a Markov chain $\mathcal{M}$ is finite if $S$ is finite.

A run is an infinite sequence of states $\rho = s_0 s_1 \dots \in S^\omega$ such that $s_0$ is a possible initial state and $s_{i+1}$ is a possible successor of $s_i$ for all $i \geq 0$. Let $\pi$ be a (finite) prefix $s_0 \dots s_n$ of a run and write $P(\pi)$ for the value $\iota_{\mathit{init}}(s_0) \prod_{i=1}^n P(s_{i-1})(s_i)$. We define the cylinder of $\pi$, written $\mathrm{Cyl}(\pi)$, as the set of all runs that have $\pi$ as a prefix. Towards defining the probability space of the Markov chain, consider the smallest $\sigma$-algebra that contains all $\mathrm{Cyl}(\pi)$, for all run prefixes $\pi$. By Carath\'eodory's extension theorem, there is a unique probability measure $\Pr$ on the $\sigma$-algebra such that for all run prefixes $\pi$:
\(
    \Pr(\mathrm{Cyl}(\pi)) = P(\pi).
\)
Hence, the expected value of a (measurable) function $f$ from runs to real numbers, written $\mathrm{E}[f]$, is also well defined via Lebesgue integrals. For example, the expected total reward, though possibly infinite, is well defined.

Write $\Diamond A$ to denote the set of 
runs that contain 
a state from $A \subseteq S$. Then, $\Pr(\Diamond A)$ is the probability that the chain eventually reaches 
a state from $A$.

\subsection{Stochastic Counter Machines}

A stochastic counter machine (or SCM) in dimension $d$ is a tuple $(Q, T)$ where $Q$ is a finite set of control states and $T \subseteq Q  \times \mathbb{Q}^d \to \mathrm{Dist}(\mathbb{Q}^d \times Q)$ is a transition relation with \emph{guards} and a distribution over successors and \emph{updates}. Intuitively, each transition is guarded by a linear constraint on the values of the $d$ counters and an affine update for the counters are sampled from a distribution along with a successor.
Concretely, the transition relation $T$ 
is given as a set of tuples $(q,g,u,q',p)$ where: $q,q' \in Q$, $g$ is a matrix-vector pair $\vec{A} \in \mathbb{Q}^{n \times d},\vec{b} \in \mathbb{Q}^n$, $u$ is also a matrix-vector pair $\vec{U} \in \mathbb{Q}^{d \times d}, \vec{r} \in \mathbb{Q}^d$, $p \in [0,1]$.

The transition is \emph{enabled} from a configuration $q(\vec{c}) \in Q\times \mathbb{N}^d$ if and 
only if $\vec{A}\vec{c} \leq \vec{b}$. Further, the successor configuration is $q'(\vec{c'})$ with probability $p$ iff $\vec{c'} = 
\vec{U}\vec{c} + \vec{r}$. 
We write $\mathrm{Sol}(g)$ to denote the set of natural solutions to 
$g(\vec{c})$ and  $\mathrm{Sol}(u)$ for the set of natural solutions to $u(\vec{c},\vec{c'})$ where both $\vec{c}$ and $\vec{c'}$ are interpreted as indeterminates. Clearly, $\vec{c} \in \mathrm{Sol}(g)$ whenever a configuration with counter values $\vec{c}$ enables a transition whose guard is $g$ and $(\vec{c},\vec{c'}) \in \mathrm{Sol}(u)$ if additionally $\vec{c'}$ corresponds to the counter values after the update of the transition.

We will add constraints to make sure the machine is deterministic (with respect to the guards) and that a proper distribution is defined: 
\begin{enumerate}
    \item For all transitions $t_1 =(q,g_1,u_1,q'_1,p_1)$ and $t_2 = (q,g_2,u_2,q'_2,p_2)$ we have that
    \(
        \mathrm{Sol}(g_1) \cap \mathrm{Sol}(g_2) \neq \emptyset
    \)
    implies $t_1 = t_2$.
    \item For all $q$ and all transitions $t_i =(q,\cdot,\cdot,\cdot,p_i)$ we have that $\sum p_i = 1$.
\end{enumerate}

\paragraph{Runs and Probabilities.} We define a run as a sequence
\( \rho = q_0(\vec{c}_0)q_1(\vec{c}_1)\dots q_n(\vec{c}_n)\)  of configurations s.t. there exist transitions $(q_i, g_i, u_i, q_{i+1}, p_i)$ 
with:
$\vec{c_i} \in \mathrm{Sol}(g_i)$,
$(\vec{c_i},\vec{c_{i+1}}) \in  \mathrm{Sol}(u_i)$, and
$p_i > 0$
for all $0\leq i < n$. 

Now, the probability of following run $\rho$ is given by $\Pr(\rho) = \prod_{i=0}^n p_i$.
We say a configuration $\chi'$ is reachable from $\chi$ iff there exists a run $\rho$ from $\chi$ to $\chi'$, which we denote with $\chi \xrightarrow{*} \chi'$. We now define the probability of reaching $\chi'$ from $\chi$, or $\Pr(\chi \xrightarrow{*} \chi')$, as $\sum_{\rho \in R} \Pr(\rho)$, where $R$ is the set of all runs from $\chi$ to $\chi'$ that contain $c_t$ exactly once as last configuration.

\paragraph{Relation with Markov chains.}
We now state the relation between Stochastic Counter Machines and Deterministic Markov Chains.
\begin{theorem}
    Let $\mathcal{A}$ be an SCM and let $\chi,\chi'$ be configurations in $\mathcal{A}$. Then, there is a Markov chain $\mathcal{M}$ with a state $s$ in such that $\Pr(\chi \xrightarrow{*} \chi')$ iff $\Pr(\Diamond  s)$. 
\end{theorem}

This can easily be shown by constructing $\mathcal{M}$, where we create a state for each possible configuration in $Q \times \mathbb{Q}^n$. For any two configurations $q_1(\vec{c}_1)$ and $q_1(\vec{c}_2)$ there is a transition in $\mathcal{M}$ with probability $p$ iff there exists a transition $(q_1, g, u, q_2, p)$ in $\mathcal{A}$ s.t. $\vec{c}_1 \in \mathrm{Sol}(g)$ and $(\vec{c_1}, \vec{c_2}) \in \mathrm{Sol}(u)$. 
It follows that the Markov chain is finite if the counters in the counter machine are bounded.

\begin{lemma}
    Let $\mathcal{A}$ be a stochastic counter machine and $\chi$ a fixed initial configuration. If there exists $u, l \in \mathbb{Q}$ such that $l \leq \lVert \vec{v} \rVert_1 \leq u$ for every reachable configuration $q(\vec{v})$ then the corresponding Markov chain is finite.   
\end{lemma}

%% file: sections/CompartmentToSCM.tex
\section{Towards Stochastic Counter Machines}

 \label{sec:bingadget}
Since the compartmental models use binomial distributions to represent the rate at which individuals move between compartments, we need a way to encode binomial distributions in stochastic counter machines. Hence, we introduce two gadgets that we 
later use in the construction of the full 
machine.

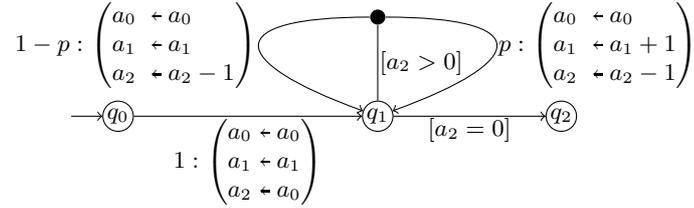
\begin{figure}
    \centering
    \begin{tikzpicture}[initial text={},inner sep=0.5pt,minimum size=0mm,every
      node/.style={font=\small},every state/.style={minimum size=0mm}]
        \node[state, initial](c0) {$q_0$} ;
        \node[state,right=3cm of c0](c1){$q_1$};
        \node[state,right=2cm of c1](c2){$q_2$};

        \node[above= of c1, circle, minimum size=0.2cm, fill] (dot){};
        \path[-] (c1)  edge  node [right] {$[a_2 > 0]$} (dot);
        \draw [->] (dot) to[out=180, in=160, looseness=4, edge node={node [left] 
            {$1-p: \begin{pmatrix*}[l] a_0 & {} \vgets a_0 \\ a_1 & {} \vgets a_1 \\ a_2 & {} \vgets a_2-1 \end{pmatrix*}$}}] (c1);

        \draw [->] (dot) to[out=0, in=20, looseness=4, edge node={node [right] {$p: \begin{pmatrix*}[l] a_0 & {} \vgets a_0\\ a_1 & {} \vgets a_1 + 1\\ a_2 & {} \vgets a_2 - 1\end{pmatrix*}$}}] (c1);
        
        \path[->,auto]
        (c0) edge node[below,]{$1:\begin{pmatrix} a_0 \vgets a_0 \\ a_1 \vgets a_1 \\ a_2 \vgets a_0 \end{pmatrix}$} (c1)
        (c1) edge node[below]{$[a_2 = 0]$} (c2);
    \end{tikzpicture}
    \caption{The standard binomial gadget: solidly filled circles represent a distribution and arrows from them lead to all possible successors; guards are shown as systems of linear constraints in square brackets on the edge between state and distribution; updates, within braces on edges from distributions to states}
    \label{fig:bingad1}
\end{figure}

The first gadget, 
\autoref{fig:bingad1}, encodes a 
binomial distribution with success rate $p$. Intuitively, we 
simulate the Bernoulli trial for each member of the source compartment. This we do by using a counter to keep track of how many members are left and a self loop that decrements the member counter increments the amount of members that follow the transition with a probability of $p$. 

%

\begin{lemma} \label{lem:bingadget1}
    Let $B(n, p)$ be a binomial distribution. Then, we can construct a SCM $\mathcal{M}$ of dimension $3$ s.t. $\Pr(k; n, p) = \Pr(q_0(n, 0, 0) \xrightarrow{*} q_2(n, k, 0))$. 
\end{lemma}

The second gadget 
allows us to model a binomial distribution where the success rate is a function of a set of variables $X= \{ x_0, x_2, \dots, x_{n-1}\}$ which we will use to represent the numbers of individuals per compartment. More precisely, the function is $1 - \prod_{i = 0}^{n-1} p_i^{x_i}$, where $p_i \in [0, 1]$ is a constant for $0 \leq i < n$. 

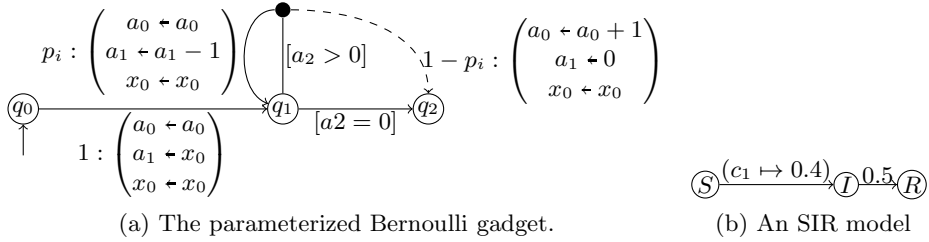
\begin{figure}
    \centering
    \subfloat[The parameterized Bernoulli gadget.
    \label{fig:gad3}]{%
    \begin{tikzpicture}[initial text={},inner sep=0.5pt,minimum size=0mm,every
      node/.style={font=\small},every state/.style={minimum size=0mm}]
        \node[state, initial below](q0){$q_0$};
        \node[state, right=3cm of q0] (q1){$q_1$};
        \node[state, right=1.5cm of q1](q2){$q_2$};
        \node[above= of q1, circle, minimum size=0.2cm, fill] (dot){};

      \draw [->] (q0)  edge node [below] {$1:\begin{pmatrix} a_0 \vgets a_0 \\ a_1 \vgets x_0 \\ x_0 \vgets x_0  \end{pmatrix}$} (q1);
      \draw [-] (q1) edge node [right] {$[a_2 >0]$} (dot);
      \draw [->] (dot) to[out=180, in=160, looseness=1, edge node={node [left] {$p_i:\begin{pmatrix} a_0 \vgets a_0 \\ a_1 \vgets a_1 -1 \\ x_0 \vgets x_0  \end{pmatrix}$}}] (c1);
      \draw [->, dashed] (dot) to[out=0, in=90,  edge node={node [pos=0.8,above, right] {$1-p_i:\begin{pmatrix} a_0 \vgets a_0 + 1 \\ a_1 \vgets 0 \\ x_0 \vgets x_0  \end{pmatrix}$}}] (q2);
      \draw [->] (q1)  edge node [below] {$[a2 = 0] $} (q2);
    
    \end{tikzpicture}
    }\hfill    
    \subfloat[An SIR model\label{fig:SIR_CMS}]{%
    \begin{tikzpicture}[initial text={},inner sep=0.5pt,minimum size=0mm,every
      node/.style={font=\small},every state/.style={minimum size=0mm}]
        \node[state](c0){$S$};
        \node[state,right=1.5cm of c0](c1){$I$};
        \node[state,right=0.5cm of c1](c2){$R$};
        
        \path[->,auto]
        (c0) edge node{$(c_1 \mapsto 0.4)$} (c1)
        (c1) edge node[above]{$0.5$} (c2);
    \end{tikzpicture}
    }
    \caption{An SCM on the left; A simple compartmental model on the right}
\end{figure}

The parameterized Bernoulli gadget, in \autoref{fig:gad3}, shows how we can do a single Bernoulli trial with single counter value $x_0$. We later elaborate on how to extend this to an arbitrary number of counters. The idea of the gadget is that, in order to not increase counter $a_0$, we have to follow a transition with probability $p_i$ consecutively exactly $x_0$ times.

\begin{lemma}\label{lem:gadget2}
    Let $p_0, p_2, \dots, p_{n-1} \in [0, 1]$ be constants. Then, we can construct a stochastic counter machine  $\mathcal{A}$ in dimension $n + 2$ such that the following hold for all $\vec{c} = (c_0,\dots,c_{n_1}) \in \mathbb{N}^n$.
    \begin{itemize}
        \item $\Pr(q_0(k, 0, \vec{c}) \xrightarrow{*} q_2(k+1, 0, \vec{c})) = 1 - \prod_{i = 0}^{n-1} p_i^{c_i}$
        
        \item $\Pr(q_0(k, 0, \vec{c}) \xrightarrow{*} q_2(k, 0, \vec{c})) = \prod_{i = 0}^{n-1} p_i^{c_i}$.
    \end{itemize} 
\end{lemma}

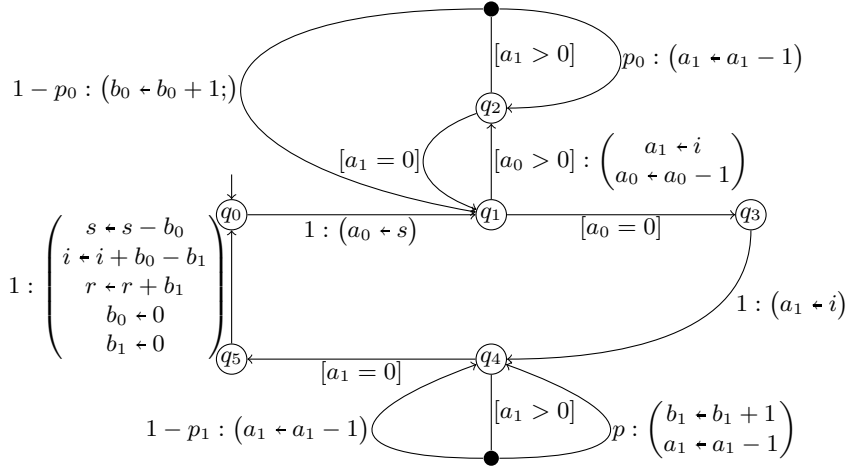
\begin{figure}
    \centering
    \begin{tikzpicture}[scale = 0.8, initial text={},inner sep=0.5pt,minimum size=0mm,every
      node/.style={font=\small},every state/.style={minimum size=0mm}]
        \node[state, initial above](q0){$q_0$};
        \node[state, right=3cm of q0] (q1){$q_1$};
        \node[state, right=3cm of q1](q3){$q_3$};
        \node[state, above= of q1] (q2) {$q_2$};
        \node[above= of q2, circle, minimum size=0.2cm, fill] (dot1){};

        \node[state, below= 1.5cm of q1] (q4) {$q_4$};
        \node[state, below= 1.5cm of q0] (q5) {$q_5$};
        \node[below= of q4, circle, minimum size=0.2cm, fill] (dot2){};

        \draw [->] (q0)  edge node [below] {$1:\begin{pmatrix} a_0 \vgets s  \end{pmatrix}$} (q1);
        \draw [->] (q1)  edge node [right] {$[a_0 > 0] : \begin{pmatrix}
            a_1 \vgets i \\ a_0 \vgets a_0 -1
        \end{pmatrix}$} (q2);
        \draw [->] (q1)  edge node [below] {$[a_0 = 0]$} (q3);

        \draw[-] (q2) edge node [right] {$[a_1>0]$} (dot1);

        \draw [->] (q2) to[out=200, in=160, looseness=2, edge node={node [left] {$[a_1 = 0]$}}] (q1);

        \draw [->] (dot1) to[out=180, in=170, looseness=4, edge node={node [left] {$1-p_0:\begin{pmatrix}
            b_0 \vgets b_0 + 1;
        \end{pmatrix}$}}] (q1);

        \draw [->] (dot1) to[out=0, in=0, looseness=4, edge node={node [right] {$p_0:\begin{pmatrix}
            a_1 \vgets a_1 - 1
        \end{pmatrix}$}}] (q2);

        \draw [->] (q3) to[out=270, in=0, looseness=1, edge node={node [pos=0.3,right=0.2cm] {$1:\begin{pmatrix}
            a_1 \vgets i
        \end{pmatrix}$}}] (q4);

        \draw [->] (q4)  edge node [below] {$[a_1 = 0] $} (q5);
        
        \draw [->] (q5)  edge node [left] {$ 1: \begin{pmatrix}
            s \vgets s - b_0 \\ 
            i \vgets i + b_0 -b_1 \\ 
            r \vgets r + b_1 \\
            b_0 \vgets 0 \\
            b_1 \vgets 0 \\
        \end{pmatrix}$} (q0);

        \draw[-] (q4) edge node [right] {$[a_1 > 0]$} (dot2);

        \draw [->] (dot2) to[out=0, in=-20, looseness=4, edge node={node [right] {$p:\begin{pmatrix}
            b_1 \vgets b_1 + 1\\
            a_1 \vgets a_1 - 1
        \end{pmatrix}$}}] (q4);

        \draw [->] (dot2) to[out=180, in=200, looseness=4, edge node={node [left] {$1-p_1:\begin{pmatrix}
            a_1 \vgets a_1 - 1
        \end{pmatrix}$}}] (q4);
    \end{tikzpicture}
    \caption{The SCM of the SIR model, where $p_0 = 1-\exp(0.4)$ and $p_1 = 1-\exp(0.5)$.}
    \label{fig:scmsir}
\end{figure}

Now we can plug in the parameterized Bernoulli gadget into the binomial gadget
to obtain a parameterized Binomial gadget. 
We give the encoding of a simple SIR-model as studied in
\cite{alarcon2023computation}, and reproduced in \autoref{fig:SIR_CMS}. The
model uses 7 counters, to which we will refer as $s, i, r, a_0, a_1, b_0, b_1$.
The full counter machine can be seen in
\autoref{fig:scmsir}.
We can see that $b_0 \sim B(s, 1-\exp(0.4)^i)$ and $b_1 \sim B(i,
1-\exp(0.5))$. Hence, going once from $q_0$ to itself simulates exactly one
timestep in the compartmental model. 

If we would be interested in the expected
time until the end of a pandemic, we can encode it by adding rewards to the
associated Markov chain. More precisely, the reward function should be $+1$
for each transition that simulates the transition from $q_0$ to $q_1$ and $0$
for all other transitions.  

\begin{theorem}
    For every compartmental model $\mathcal{C}$ we can construct an SCM $\mathcal{A}$ with a reward function such that: (1) There is an injective mapping $\mu$ from states of the $\mathcal{C}$ to configurations of $\mathcal{A}$; (2) For all pairs $(p,q)$ of states of $\mathcal{C}$ we have $\Pr(p \rightarrow q)$ equals $\Pr(\mu(p) \xrightarrow{*} \mu(q))$ in $\mathcal{A}$; (3) Moreover, the expected time to reach $q$ from $p$ in $\mathcal{C}$ is the expected total sum before reaching $q$ from $p$ in $\mathcal{A}$.
\end{theorem}

%% file: sections/Experiments.tex
\section{Experimental Results}\label{sec:experiments}
We implemented \emph{Inform}, a tool that translates compartmental models into
stochastic counter machines. 
Concretely, \emph{Inform} translates them into a PRISM file.
%
Using it and the \emph{Storm} model checker~\cite{HenselJKQV22}, we can verify properties such as the expected end
of the epidemic. 
Since the COVID-19 model is huge, we also
compare Storm with a statistical model checker,
\emph{Modest} \cite{HartmannsH14}. The latter 
usually scale better yet provide 
confidence intervals
instead of exact probability values.

All experiments were ran on a cluster where each node has
an Intel(R) Xeon(R) Platinum 8168 CPU @ 2.70GHz with 64GiB of memory and no
GPU.
%
%
We ran Storm from the \emph{movesrwth/storm:stable} docker container. Storm's version was 1.7.1. For Modest, we used version was v3.1.237-g2f62162c7.

For the experiments we used two models. The first one is the simple SIR model with three compartments based on the model studied in \cite{alarcon2023computation} (cf. \cref{sec:simplesirmodel}).
The second model is a simplified COVID-19 model based on
\cite{ABRAMS2021100449} (cf. \cref{sec:covidmodelspec}). There, we consider different populations partitioned as $(S,I_a,I_m,I_s)$ where $S$ stands for susceptible; $I_a$, asymptomatic infectious individuals; $I_m$, mild cases; and $I_s$, severe cases. 
We denote time-outs with TO: we stopped the computation after 1 hour (for small populations); memory-outs with MO: the program was terminated because it ran out of memory.

\paragraph*{Parameter Tuning for Storm}
In \Cref{tab:sirtractable}, we ran Storm on the SIR model to determine the best parameter values. From local tests, we found that the engine parameter and the binary decision diagram library had the biggest impact on run-times. We then compared the sparse engine, the dd engine with both Sylvan and BuDDy, and the dd-to-sparse engine with both Sylvan and BuDDy.
The most favourable setting is the sparse engine. Even though decision diagrams should provide a short representation, it seems that it does not outperform the sparse engine. 
However, 
Storm starts to run out of memory for 
small populations. 

\input{tables/tableSIRStormParams}

\paragraph*{Scalability of Storm and Modest}
As mentioned before, the COVID-19 model is huge. The state space
encodes all possible partitionings of the population into 10 compartments.
%
Consider \cref{tab:modeststormperformance}. 
For both the population increase (PopInc) as well as the end of epidemic (EoE) property, we see Storm running out of memory for populations of 10. The one-shot (OS) property performs really well in Storm. This may be because our fomalization of
the OS property only checks the first time individuals change compartment. 
Storm seems to be taking this into account when building the state-space. 
Modest performs significantly better than Storm for the PopInc property. Moreover, the size of the population has a much smaller impact on the run-time compared to Storm. However, for the EoE property, we see that Modest struggles. In order to not time out for small instances, the width of the confidence interval of Modest was set to $0.9$ and, even in this case, run-times were significantly higher than for the PopInc property.

\input{tables/StormModestPerformance}

\begin{figure}[t]
    \centering
    \begin{tikzpicture}[scale=0.6]
    \begin{axis}[%
        xlabel = {$(S, 1, 1, 1)$ population},
        ylabel = {seconds},
        scatter/classes={%
        a={draw=black}}]
        \addplot[scatter,only marks,%
            scatter src=explicit symbolic]%
        table[meta=label] {
            x y label
            10  6.1 a  \\
            20  9.3 a \\
            40  11.7 a \\
            80  12.1 a \\
            160  25.2 a \\
            320  27.7 a \\
            640  68.1 a \\
            1280  74.7 a \\
            2560  257.1 a \\
            5120  504.0 a \\
            10240 689.8 a \\
            20480 1869.1 a \\
            40960 2774.2 a \\
            81920 6420.6 a \\
            163840 8882.6 a \\
            327680 29609.8 a \\
        };
    \end{axis}
\end{tikzpicture}
    \caption{The a plot of Modest run-times vs. large populations for PopInc.}
    \label{fig:popincultimate}
\end{figure}
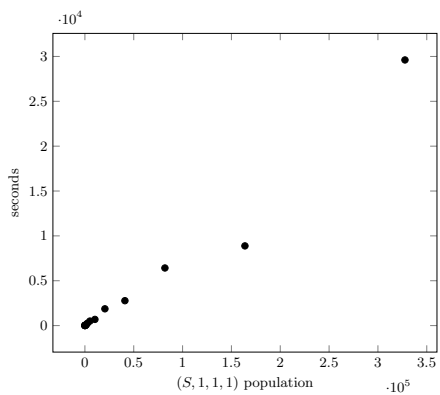

Finally, for \Cref{fig:popincultimate}, we used Modest to analyze the COVID-19 model with realistic populations. The used width here was the default $0.01$ and the populations were $(S, 1, 1, 1)$ with $S$ is increasing. We observe that the run-time of Modest grows almost linearly with respect to $S$.

%% file: tables/tableSIRStormParams.tex
\begin{table}[t]
    \centering
    \caption{Scalability of Storm on EoE property for SIR model.}
    \begin{tabular}{|c|c|c|c|c|c|}
        \hline
        &&& dd-to-sparse & & dd-to-sparse\\
        population & sparse &  dd \& BuDDy & BuDDy &  dd \& Sylvan & Sylvan  \\
        \hline
        (5, 5, 0) & 0.074s & 402.931s   & 68.813s    & 14.552s    & 0.276s    \\
        (10, 10, 0)& 1.524s & 1682.572s  & 175.385s   & 312.827s   & 7.408s    \\
        (15, 15, 0)& 18.908s & 4446.139s  & 467.44s    & 1984.448s  & 61.682s   \\
        (20, 20, 0)& 118.889s & 15285.464s & 1796.469s  & 9194.601s  & 412.937s  \\
        (25, 25, 0)& 521.013s & MO         & 8483.432s  & 31162.389s & 2008.203s \\
        (30, 30, 0)& 1825.479s & MO         & 21577.803s & MO         & 7932.179s \\
        (35, 35, 0)& MO & MO         & MO         & MO         & MO        \\
        
        \hline
    \end{tabular}
    \label{tab:sirtractable}
\end{table}

%% file: tables/StormModestPerformance.tex
\begin{table}[t]
    \centering
    \caption{Performance of Storm compared to Modest on the COVID-19 model. For Storm, we used default parameters and the sparse engine. For the EoE property, Modest was run with max run length 0 and width 0.9; for the others, with max run length 0 and width 0.01.}
    
    \begin{tabular}{|c|c|c|c|}
        \hline
        Property & Population & Storm runtime & Modest runtime \\
        \hline
        PopInc & (2, 1, 1, 1) & 2.916s &  3.4s \\
         & (3, 1, 1, 1) & 21.201s &  6.9s \\
         & (4, 1, 1, 1) & 297.23s & 5.9s  \\
         & (5, 1, 1, 1) & 2352.066s & 5.7s  \\
         & (6, 1, 1, 1) & 14756.769s & 4.5s  \\
         & (7, 1, 1, 1) & MO & 4.3s  \\
         \hline
         EoE & (2, 1, 1, 1) & 3.400s & 1023.5s  \\
         & (3, 1, 1, 1) & 27.314s &  997.8s \\
         & (4, 1, 1, 1) & 570.862s &  1069.3s \\
         & (5, 1, 1, 1) & 5325.083s &  1048.3s \\
         & (6, 1, 1, 1) & 42751.95 &  1039.0s \\
         & (7, 1, 1, 1) & MO &  1080.8s \\
         \hline
      OS & (2, 1, 1, 1) & 0.123s & 0.2s  \\
         & (3, 1, 1, 1) & 0.125s & 0.1s  \\
         & (4, 1, 1, 1) & 0.141s & 0.2s  \\
         & (5, 1, 1, 1) & 0.186s & 0.1s  \\
         & (6, 1, 1, 1) & 0.195s & 0.1s  \\
         & (7, 1, 1, 1) & 0.223s & 0.1s  \\
         & (8, 1, 1, 1) & 0.244s & 0.2s  \\
         & (9, 1, 1, 1) & 0.274s & 0.1s  \\
         & (10, 1, 1, 1) & 0.306s & 0.2s  \\
         \hline
    \end{tabular}
    \label{tab:modeststormperformance}
\end{table}

%% file: sections/conclusion.tex
\section{Conclusion}
We have described a simple translation from stochastic compartmental models to Markov chains. The translation has been implemented in a tool that should help epidemiologists analyze new models modified to capture the dynamics of different pathogens.
From the experiments, we can conclude that state-of-the-art probabilistic model checkers are not (yet) powerful enough to deal with epidemiological models like the one proposed in \cite{ABRAMS2021100449}. Indeed, while Modest is quite capable of handling simple probabilistic queries for realistic populations, it still seems to struggle with quantitative queries such as the expected end of epidemic property. In this direction, more research is needed in finding good abstractions that, for example, reduce the state space.
Recent work shows promising efficient techniques for smarter treatment of conditional expected values (which is needed for the EoE property) \cite{perez23atva}. Another interesting direction consists in leveraging our encoding to enable reinforcement learning (see, e.g.~\cite{DBLP:conf/setta/GrossJJP22}) of intervention policies.

%% file: sections/appendix.tex
\section{Properties in Prism Format}
Below we state a table with the properties that were used in the experiment section.

\begin{table}[]
    \centering
    \caption{The properties of interest stated in PRISM-style probabilistic computation tree logic and in terms of the Belgian COVID-19 model.}
    \begin{tabular}{|c|c|c|}
        \hline
        Question & ID & Property  \\
        \hline
        1.  & popinc & P=? [$(q < 31)$ U $(c1 + c2 + c3 + c4 + c5 = 0)$] \\
        2.  & OS & P=? [$(c0 >= c0\_init)$ U $(q=30)\&(fr0 = c0\_init)$] \\
        3.  & EoE & R\{time\_step\}=? [F $(c1+c2+c3+c4+c5 = 0)$]"\\
        \hline
        \end{tabular}
    
    \label{tab:properties}
\end{table}

\section{Missing Proofs of \Cref{sec:bingadget}}

Here we state the missing proof of \cref{lem:bingadget1}:

\begin{proof}
    First, note that  $q_2$ can only be reached if the self loop in $q_1$ has been taken exactly $n$ times. This is because the guard on the transition to reach $q_2$ requires the temporary variable in $a_2$ to be $0$ and each time the self loop is taken, we decrease $a_2$ with one. 

    Now, to reach $q_2(n, k, 0)$, we need to have taken the right transition from $q_1$ exactly $k$ times. Hence the probability of reaching $q_2(n, k, 0)$ with some run is exactly $p^k(1-p)^{(n-k)}$. Now, since the order doesn't matter for taking the right transition transition, there are exactly $n \choose k$ possible runs from $q_0$ to $q_2$ that do not contain $q_2$. Hence, the probability of reaching $q_2(n, k, 0)$ from $q_0(n, 0, 0)$ is 
    \[
        {n \choose k} p^k(1-p)^{n-k} = \Pr(k; n, p).
    \]
    This concludes the proof.\qed
\end{proof}

Here we state the missing proof of \cref{lem:gadget2}:
\begin{proof}
    Consider the gadget in \autoref{fig:gad3} and let $c \in \mathbb{N}$ be arbitrary. We refer to the first two counters as $a_0$ and $a_1$, the other counters will be referred to with the parameter they represent. It is clear to see that $a_0$ can only be incremented if we take the dashed transition. We now show that the probability of reaching $q_2$ without incrementing $a_0$ is $p^c$ if the value of $x$ is $c$. Indeed the only way to reach $q_2$ with no increment is when $a_2 = 0$ in $q_1$, which can only happen when we followed $c$ times the self loop on $q_1$.  The probability of that run is $p^{c}$, and hence, the probability of incrementing $a_1$ is $1-p^{c}$.

    Now we can easily chain these gadgets for each $x_i \in X$, where we connect all dashed transitions to the final state in the chain. Then, for any $\vec{c} \in \mathbb{N}^n$, the probability of the run that does not increment $a_0$ (and follows only full drawn transitions) is $\prod_{i=0}^{n-1} p_i^{c_i}$ and the probability of following at least one dashed transition is $1-\prod_{i=0}^{n-1} p_i^{c_i}$. \qed
\end{proof}

\section{The Simple SIR Model} \label{sec:simplesirmodel}
The simple SIR model described in \cite{alarcon2023computation} uses three compartments: Susceptible, Infectious, and Removed. Although inform is able to compute this model from the CMS file given in 

The infection rate $p = 0.7408182206817179$ and the recovery rate $r = 0.3934693402873666$ are the exact probabilities used. 

\begin{figure}[htbp]
\begin{tabular}{p{0.5\textwidth}p{0.5\textwidth}}
    \begin{minipage}{.5\textwidth}
    \centering
    \begin{tikzpicture}[initial text={},inner sep=0.5pt,minimum size=0mm,every
      node/.style={font=\small},every state/.style={minimum size=0mm}]
        \node[state](c0){$S$};
        \node[state,right= of c0](c1){$I$};
        \node[state,below= of c1](c2){$R$};
        
        \path[->,auto]
        (c0) edge node[above]{$(p, I)$} (c1)
        (c1) edge node{$(r)$} (c2);
        
    \end{tikzpicture}
    
    \label{fig:fig1}
    \end{minipage}
    &
    \begin{minipage}{.5\textwidth}
\begin{lstlisting}
-meta-
pop 10
h 1
comps 3
final 1
-trans-
0 1 [0.2 1] 
1 2 [0.3 _] 
\end{lstlisting}
\end{minipage}
\end{tabular}
\caption{On the left the graph representation of the SIR model, on the right the cms file of the SIR model.}
\end{figure}
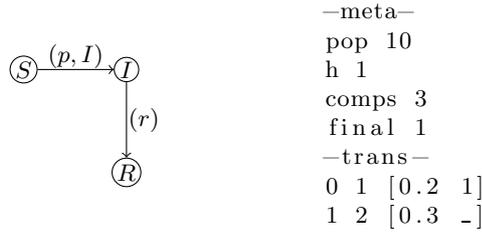

\section{Values for the COVID-19 Model}\label{sec:covidmodelspec}
In this section we provide the values we used to build the model for the early stages of COVID. The original model in \cite{ABRAMS2021100449} featured multiple age groups, but our encoding is not yet able to handle models of that size. Hence, we focussed on the agegroup [20 - 30] or age group 3. In the table, we show the associated transition, the computation, and the actual value of the success probability.

Since we're dealing with real numbers, we have to use some finite representation. Hence, we list all values in the precision used in the model. For the Euler constant we used the value: $e = 2.7182818284$, as for the timestep, we used $h=0.04166666666666666$.

\begin{table}[]
    \centering
    \begin{tabular}{|c|c|c|c|}
    \hline
        Transition & Parameter & Parameter Value & Probability \\
        \hline
        $E\rightarrow I_{presym}$ & $\gamma$ & $0.729$ & 0.0299183253\\ 
        $I_{presym} \rightarrow I_{asym}$ & $(p)\theta$ & $(0.84)0.475 = 0.399$ & 0.0164875673
\\
        $I_{presym} \rightarrow I_{mild}$ & $(1-p)\theta$ & $(1-0.84)0.475 = 0.076$ & 0.0031616581\\

        $I_{asym} \rightarrow R$ & $\delta_1$ &0.240 &0.0099501663\\
        
        $I_{mild} \rightarrow R$ & $\delta_2 = \phi_0\delta^*_2$&$ 0.984*0.756 =0.743904$ & 0.0305205490\\

        $I_{mild} \rightarrow I_{sev}$ & $\psi = (1 - \phi_0)\delta^*_2$&$ (1-0.984)*0.756 =0.012096$ & 0.0005038730\\

        $I_{sev} \rightarrow I_{hosp}$ & $\phi_1\omega$ & 0.75*0.099 = 0.07425 & 0.0030889693\\
        $I_{sev} \rightarrow I_{icu}$ & $(1-\phi_1)\omega$ & 0.25*0.099 = 0.02475 & 0.0010307184\\
        
        $I_{hosp} \rightarrow R$ & $\delta_3 = (1-\mu)\delta^*_3$ & $(1-0.005)0.185 = 0.184075 $ & 0.0076404538\\
        $I_{hosp} \rightarrow D$ & $\tau_1 = \mu \delta^*_3$ & 0.000925 & 0.0000385409 \\
        $I_{icu} \rightarrow R$ & $\delta_4 = \delta_3$ & 0.184075& 0.0076404538\\
        $I_{icu} \rightarrow D$ & $\tau_2 = \tau_1$ & 0.000925& 0.0000385409\\
        \hline
    
    \end{tabular}
    \caption{Caption}
    \label{tab:my_label}
\end{table}

Finally, we have to define the success rate for the transition $S\rightarrow E$. This is computed as follows:
\[
   \lambda =  q C_{asym} (I_{pre} + I_{asym}) + q C_{sym} (I_{mild} + I_{sev}).
\]

We used the values $q=0.051$, $C_{asym} = 6.06$, and $C_{sym} = 1.52$. Hence, the probability of success is $1 - exp(-h 0.30906 (I_{pre} + I_{asym}) + 0.07752 (I_{mild} + I_{sev}))$ or 
\[
    1-( 0.0127949398^{(I_{pre} + I_{asym})}0.0032247892^{(I_{mild} + I_{sev})}.
\]

\begin{figure}
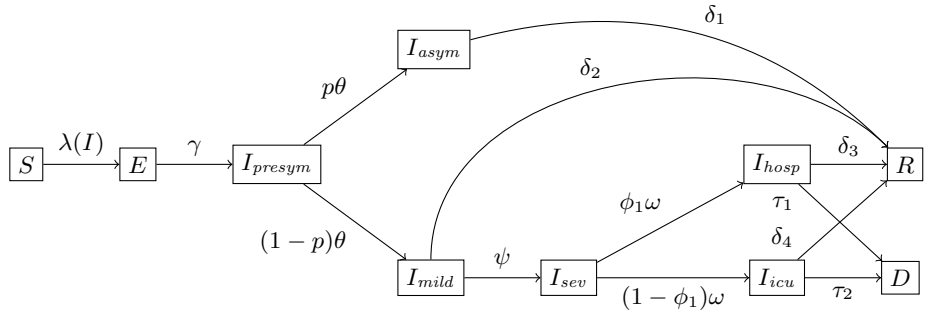

    \centering
    \include{images/sircovid}
    \caption{The full compartmental model for the early stages of COVID in Belgium, as presented in \cite{ABRAMS2021100449}}.
    \label{fig:enter-label}
\end{figure}

\section{Modest Precision}

To check that Modest computes good approximate values, we compare the outcome of the tools for small populations in \cref{tab:modestprecision}. For a width of at most $0.01$, the values are reasonably good.

\input{tables/StormModestPrecision}

Note that Modest times out on the $(3, 0, 0, 0)$ population for the one-shot property. This is because formulation of property checks whether everyone gets infected when at least one person gets infected. This never occurs in this edge case and we see that modest is not able to detect this, while storm is. Additionally, the probability of a "one-shot infection" is very high, while we stated earlier that it should be very low. This is due to population size. Note that this size models a single household and from experience we can conclude that this is realistic behavior.  

\section{Full Table 2}

In \cref{sec:experiments}, we give a short version of \cref{tab:modeststormperformance}. Here we give the full table.

\begin{table}[]
    \centering
    \caption{Performance of Storm compared to Modest on the COVID model. For Storm, we used default parameters and the sparse engine. For the eoe property, modest was run with max run lenght 0 and width 0.9. For the other properties, Modest was run with max run length 0 and width 0.01. }
    \begin{tabular}{|c|c|c|c|}
        \hline
        property & population & Storm runtime & Modest runtime \\
        \hline
  popinc & (2, 1, 1, 1) & 2.916 &  3.4s \\
         & (3, 1, 1, 1) & 21.201 &  6.9s \\
         & (4, 1, 1, 1) & 297.23s & 5.9s  \\
         & (5, 1, 1, 1) & 2352.066s & 5.7s  \\
         & (6, 1, 1, 1) & 14756.769s & 4.5s  \\
         & (7, 1, 1, 1) & MO & 4.3s  \\
         & (8, 1, 1, 1) & MO & 5.3s  \\
         & (9, 1, 1, 1) & MO & 3.5s  \\
         & (10, 1, 1, 1) & MO & 6.2s \\
         \hline
         EoE & (2, 1, 1, 1) & 3.400s & 1023.5s  \\
         & (3, 1, 1, 1) & 27.314s &  997.8s \\
         & (4, 1, 1, 1) & 570.862s &  1069.3s \\
         & (5, 1, 1, 1) & 5325.083s &  1048.3s \\
         & (6, 1, 1, 1) & 42751.95 &  1039.0s \\
         & (7, 1, 1, 1) & MO &  1080.8s \\
         & (8, 1, 1, 1) & MO &  1120.8s \\
         & (9, 1, 1, 1) & MO &  1124.7s \\
         & (10, 1, 1, 1) & MO &  1134.0s \\
         \hline
      OS & (2, 1, 1, 1) & 0.123s & 0.2s  \\
         & (3, 1, 1, 1) & 0.125s & 0.1s  \\
         & (4, 1, 1, 1) & 0.141s & 0.2s  \\
         & (5, 1, 1, 1) & 0.186s & 0.1s  \\
         & (6, 1, 1, 1) & 0.195s & 0.1s  \\
         & (7, 1, 1, 1) & 0.223s & 0.1s  \\
         & (8, 1, 1, 1) & 0.244s & 0.2s  \\
         & (9, 1, 1, 1) & 0.274s & 0.1s  \\
         & (10, 1, 1, 1) & 0.306s & 0.2s  \\
         \hline
    \end{tabular}
    \label{tab:odeststormperformance}
\end{table}

%% file: images/sircovid.tex
\begin{tikzpicture}
    \node[rectangle,draw](s){$S$};
    \node[rectangle,draw,right= of s](e){$E$};
    \node[rectangle,draw,right= of e](ip){$I_{\mathit{presym}}$};
    \node[rectangle,draw,above right= of ip](ia){$I_{\mathit{asym}}$};
    \node[rectangle,draw,below right= of ip](im){$I_{\mathit{mild}}$};
    \node[rectangle,draw,right= of im](is){$I_{\mathit{sev}}$};
    \node[rectangle,draw,right=2cm of is](ii){$I_{\mathit{icu}}$};
    \node[rectangle,draw,above= of ii](ih){$I_{\mathit{hosp}}$};
    \node[rectangle,draw,right= of ih](r){$R$};
    \node[rectangle,draw,right= of ii](d){$D$};
    \path[->,auto]
      (s) edge node{$\lambda(I)$} (e)
      (e) edge node{$\gamma$} (ip)
      (ip) edge node{$p\theta$} (ia)
      (ip) edge node[swap]{$(1-p)\theta$} (im)
      (ia) edge[out=15] node{$\delta_1$} (r)
      (im) edge[out=90] node{$\delta_2$} (r)
      (im) edge node{$\psi$} (is)
      (is) edge node{$\phi_1\omega$} (ih)
      (is) edge node[swap]{$(1-\phi_1)\omega$} (ii)
      (ih) edge node{$\delta_3$} (r)
      (ih) edge node[pos=0.05,swap]{$\tau_1$} (d)
      (ii) edge node[swap]{$\tau_2$} (d)
      (ii) edge node[pos=0.05]{$\delta_4$} (r)
      ;
  \end{tikzpicture}

%% file: tables/StormModestPrecision.tex
\begin{table}[]
    \centering
    \caption{The precision of Modest compared to the exact values in Storm. Note that for the EoE property, we used width 0.9 to keep the run time manageable in storm. The random number generator for Modest was set to 997.}
    \begin{tabular}{|c|c|c|c|c|}
        \hline
        property & population & Storm & Modest (width 0.01) &   Modest (width 0.001)  \\
        \hline
         popinc&(3, 0, 0, 0) & 1 & 1 & 1 \\
         &(0, 1, 1, 1) & 0.9987240195 & 1 & 0.9989641434262948  \\
         &(3, 1, 1, 0) & 0.9952634065 & 0.9971428571428571 & 0.9955811462238886\\
         &(2, 1, 1, 1) & 0.9948733242 & 0.9971428571428571 & 0.9953703703703703\\
         &(1, 4, 0, 0) & 0.9972593727 & 0.9905982905982906  &  0.9971220260936301  \\
         \hline
         EoE &(3, 0, 0, 0) & 0 & 0 & 0 \\
         &(0, 1, 1, 1) & 275.6801531 & 276.65685600648203  & TO \\
         &(3, 1, 1, 0) & 285.2523967 & 285.1537294071891 & TO \\
         &(2, 1, 1, 1) & 348.3760444 & 348.7204048428036 & TO \\
         &(1, 4, 0, 0) & 255.5465313 & 255.65503788941862 & TO \\
         \hline
         OS &(3, 0, 0, 0) & 0 & TO & TO\\
         &(0, 1, 1, 1) & 1 & 1 & 1\\
         &(3, 1, 1, 0) & 0.999876224 & 1 & 1\\
         &(2, 1, 1, 1) & 0.9999997331 & 1 & 1 \\
         &(1, 4, 0, 0) & 1  & 1 & 1\\
         \hline
    \end{tabular}
    \label{tab:modestprecision}
\end{table}